\documentclass{article}
\usepackage{amsmath,amsthm,amssymb}

\newtheorem{theorem}{Theorem}[section]
\newtheorem{corollary}[theorem]{Corollary}
\newtheorem{lemma}[theorem]{Lemma}
\newtheorem{proposition}[theorem]{Proposition}
\theoremstyle{remark}
\newtheorem{remark}[theorem]{Remark}
\theoremstyle{definition}
\newtheorem{definition}{Definition}

\newcommand\R{{\mathbb R}}
\newcommand\N{{\mathbb N}}
\newcommand\M{{\mathcal M}}
\newcommand\B{{\mathcal B}}
\newcommand\G{{\mathcal G}}
\newcommand\Ga{\Gamma}
\newcommand\ga{\gamma}
\newcommand\La{\Lambda}
\newcommand\la{\lambda}

\newcommand\1{{1 \!\! 1}}
\newcommand\X{X}

\numberwithin{equation}{section}
\allowdisplaybreaks[3]

\begin{document}

\title{Measures on two-component configuration spaces%
\thanks{I would like to thank Prof. Dr. Yuri Kondratiev for useful discussions.
This work was partially supported by DFG through SFB 701, Bielefeld
University.}}
\author{\textbf{D.L. Finkelshtein}\\
\small{Institute of Mathematics, Ukrainian National Academy of
Sciences}\\ \small{3 Tereshchenkivs'ka str, Kyiv-4, 01601, Ukraine}\\
\small{e-mail: fdl@imath.kiev.ua.}}

\date{}

\maketitle

\begin{abstract}
We study measures on the configuration spaces of two type particles.
Gibbs measures on the such spaces are described. Main properties of
corresponding relative energies densities and correlation functions
are considered. In particular, we show that a support set for the
such Gibbs measure is the set of pairs of non-intersected
configurations.
\end{abstract}

\bigskip


\bigskip


\newpage

\section{Introduction}
The study of measures and related objects on the spaces of infinite
configurations in Euclidean spaces (or, more general, $C^\infty$
manifolds) was started in sixties. In 1979, in \cite{NZ}, it was
considered several approaches to describing Gibbs measures on the
configuration spaces. Different aspects of the corresponding measure
theory were discovered in \cite{KMM}, \cite{MMW}, \cite{Pap},
\cite{Kall}, \cite{Gl1}, \cite{Gl2}, \cite{WE}, \cite{GK} and
others. For the case of marked configurations the
Dobrushin---Lanford---Ruelle (DLR) approach was considered in
\cite{KonKunSilva}, \cite{Kun01}. Nevertheless, describing of marked
Gibbs measures via integral equations (so-called,
Georgii---Nguyen---Zessin---Campbell---Mecke equations) was not
realized.

In this work we study these equations for the simplest case of the
space of marks: $\{+,-\}$. We extend approach proposed in
\cite{FK05} for this marked (two-component) system. We concentrate our
attention on the properties of the Gibbs type measures without
studying existence and uniqueness problems. One may study this using
Ruelle technique in the same way as in \cite{FK05}, which we
represent in the forthcoming paper. Another approach for proving
existence and non-uniqueness was proposed in \cite{GH}.

Let us describe the content of the work in more detail.

Preliminary constructions for the one-component case are presented
in Section~2. In Section~3 we consider main properties of a measure
on the two-component configuration spaces which is locally
absolutely continuous with respect to (w.r.t.) product of two
Poisson measures. Note that it is natural that these Poisson
measures have the same intensities since they should not be
orthogonal. This is impossible for different constant intensities
but for non-constant ones we need some additional conditions (see,
e.g., \cite{Tak}). Hence, for simplicity we consider the same Poisson
measures. One of the main results of this section is connection
between correlation functions of a measure and of their marginal
distribution. In Section~4 we describe the Gibbs measures in terms
of the so-called relative energies densities, which characterized
the energy between particle of one type and configurations of the
both types. Main properties of these densities allow us to show that
the corresponding Gibbs measure is locally absolutely continuous
w.r.t. product of Poisson measures. As a result, we may study
such measure only on the subspace of the two-component configuration
space which includes only pairs of configurations which are not
intersect. This plays important role for studying different dynamics
on the two-component configuration spaces, namely, we have useful
support set for a big class of measures (see, e.g., \cite{FKS08},
\cite{FinFilKo08}). At we end we show an example of the
pair-potentials Gibbs measure which coincides with studying in
\cite{GH}.

We don't construct in this work specifications of the Gibbs measure
and corresponding DLR approach. This may be considered analogously
to \cite{FK05} as well as it possible to show the equivalence
between such two approaches (that goes back to \cite{NZ}). All our
considerations may be extended on the case of the product of finite
number of the configuration spaces over different $C^\infty$
manifolds.

\section{Preliminaries}
Let $\X$ be a connected oriented $C^\infty$ manifold.
The configuration space $\Ga :=\Ga _{\X}$ over $\X$ is defined as the set of all locally finite subsets of
$\X$,
\begin{equation}
\Ga :=\bigl\{ \ga \subset \X \bigm| \left| \ga_\La\right| <\infty
\text{ for every compact }\La\subset \X\bigr\} ,
\end{equation}
where $\left| \cdot \right|$ denotes the cardinality of a set and
$\ga_\La := \ga \cap \La$. As usual we identify each
$\ga \in \Ga $ with the non-negative Radon measure $\sum_{x\in
\ga }\delta_x\in \mathcal{M}(\X)$, where $\delta_x$ is
the Dirac measure with unit mass at $x$,
$\sum_{x\in\varnothing}\delta_x$ is, by definition, the zero measure,
and $\mathcal{M}(\X)$ denotes the space of all
non-negative Radon measures on the Borel $\sigma$-algebra
$\mathcal{B}(\X)$. This identification allows to endow
$\Ga $ with the topology induced by the vague topology on
$\mathcal{M}(\X)$, i.e., the weakest topology on $\Ga$
with respect to which all mappings
\begin{equation*}
\Ga \ni \ga \longmapsto \langle f,\ga\rangle :=
\int_{\X}f(x)d\ga(x)=\sum_{x\in \ga }f(x),\quad f\in C_0(\X),
\end{equation*}
are continuous. Here $C_0(\X)$ denotes the set of all continuous functions
on $\X$ with compact support. We denote by $\mathcal{B}(\Ga )$ the
corresponding Borel $\sigma$-algebra on $\Ga$.

Let us now consider the space of finite configurations
\begin{equation*}
\Ga_0 := \bigsqcup_{n=0}^\infty \Ga^{(n)},
\end{equation*}
where $\Ga^{(n)} := \Ga^{(n)}_{\X} := \{ \ga\in \Ga: \vert \ga\vert
= n\}$ for $n\in \N$ and $\Ga^{(0)} := \{\varnothing\}$. For $n\in
\N$, there is a natural bijection between the space $\Ga^{(n)}$ and
the symmetrization $\widetilde{\X^n}\diagup S_n$ of the set
$\widetilde{\X^n}:= \{(x_1,...,x_n)\in \X^n: x_i\not= x_j \hbox{ if
} i\not= j\}$ under the permutation group $S_n$ over $\{1,...,n\}$
acting on $\widetilde{\X^n}$ by permuting the coordinate indexes.
This bijection induces a metrizable topology on $\Ga^{(n)}$, and we
endow $\Ga_0$ with the topology of disjoint union of topological
spaces. By $\mathcal{B}(\Ga^{(n)})$ and $\mathcal{B}(\Ga_0)$ we
denote the corresponding Borel $\sigma$-algebras on $\Ga^{(n)}$ and
$\Ga_0$, respectively.

Given a non-atomic Radon measure $\sigma$ on $(\X, \B(\X))$ with
$\sigma(\X)=\infty$, let $\la_\sigma$ be the Lebesgue-Poisson
measure on $\bigl(\Ga_0,\B(\Ga_0)\bigr)$, namely,
$$
\la_\sigma:=\sum_{n=0}^\infty \frac{1}{n!} \sigma^{(n)},
$$
where each $\sigma^{(n)}$, $n\in \N$, is the image measure on $\Gamma^{(n)}$ of
the product measure $d\sigma(x_1)...d\sigma(x_n)$ under the mapping
$\widetilde{\X^n}\ni (x_1,...,x_n)\mapsto\{x_1,...,x_n\}\in \Gamma^{(n)}$.
For $n=0$ we set $\sigma^{(0)}(\{\varnothing\}):=1$.

Let $\mathcal{B}_c(\X)$ denote the set of all bounded Borel sets in
$\X$, and for any $\La\in \mathcal{B}_c(\X)$ let $\Ga_\La :=
\{\eta\in \Ga: \eta\subset \La\}$. Evidently $\Ga_\La =
\bigsqcup_{n=0}^\infty \Ga_\La^{(n)}$, where $\Ga_\La^{(n)}:=
\Ga_\La \cap \Ga^{(n)}$ for each $n\in \N_0$, leading to a situation
similar to the one for $\Ga_0$, described above. We endow $\Ga_\La$
with the topology of the disjoint union of topological spaces and
with the corresponding Borel $\sigma$-algebra
$\mathcal{B}(\Ga_\La)$. Let $\mathbf{p}_\La:\Ga\rightarrow\Ga_\La$ be a
projection mapping: $\mathbf{p}_\La(\ga)=\ga_\La$. Then if we define Poisson
measure on $\bigl(\Ga_\La,\B(\Ga_\La)\bigr)$ as
$\pi_\sigma^\La=e^{-\sigma(\La)}\la_\sigma$ (here we understand
$\la_\sigma$ as measure on $\Ga_\La$), it is well known that there
exists a unique Poisson measure on $\bigl(\Ga,\B(\Ga)\bigr)$ such
that $ \pi_\sigma^\La=\pi_\sigma \circ \mathbf{p}_\La^{-1}$ for any
$\La\in\B_c(\X)$.
Note that $\bigl(\Ga,\B(\Ga),\pi_\sigma\bigr)$ is a projective limit
of the family $\Bigl\{\bigl(\Ga_\La,\B(\Ga_\La),
\pi_\sigma^\La\bigr) \Bigm| \La\in\B_c(\X) \Bigr\}$.

We suppose from the beginning that there exists a sequence
$\{\La_m\}_{m\in\N}\subset\B_c(\X)$ such that
$\bigcup_{m\in\N}\La_m=\X$.

\section{Measures on two-component spaces}
Let $\Gamma ^{+}=\Gamma ^{-}=\Gamma _{X}$ and $\Gamma ^{2}=\Gamma
^{+}\times \Gamma ^{-}$. We consider a topology of direct product on
$\Ga^2$. Then $\B(\Ga^2):=\B(\Ga^+)\times \B(\Ga^-)$ is the
corresponding Borel $\sigma$-algebra. We denote a class of
probability measures on $(\Ga^2,\B(\Ga^2))$ by $\M^1(\Ga^2)$.

Let us consider a projection mapping
$p_{\La^+,\La^-}:\Ga^2\rightarrow\Ga^+_{\La^+}\times\Ga^-_{\La^-}$
such that
\[
p_{\La^+,\La^-}(\ga^+,\ga^-)=\bigl(\ga^+_{\La^+},\ga^-_{\La^-}\bigr).
\]

\begin{definition}
We call a measure $\mu\in\M^1(\Ga^2)$ {\em locally absolutely
continuous} w.r.t. $\pi_\sigma\times\pi_\sigma$ if
$\mu^{\La^+,\La^-}:=\mu\circ p_{\La^+,\La^-}^{-1}$ is absolutely
continuous w.r.t. product of the Poisson measures
$\pi_\sigma^{\La^+}\times\pi_\sigma^{\La^-}$ on
$\bigl(\Ga^+_{\La^+}\times\Ga^-_{\La^-},\B(\Ga^+_{\La^+})\times\B(\Ga^-_{\La^-})\bigr)$.
\end{definition}

In the case when $\La^+=\La^-=\La$ we will write
$p_\La,\mu^\La,\Ga_\La^2$ instead of $p_{\La,\La},
\mu^{\La,\La},\Ga^+_{\La}\times\Ga^-_{\La}$ correspondingly.

\begin{proposition}\label{goodset}
For any $\mu\in\M^1(\Ga^2)$ which is locally absolutely continuous
w.r.t. $\pi_\sigma\times\pi_\sigma$ the set
\begin{equation}
\widetilde{\Ga}^2:=\Bigl\{ (\ga^+,\ga^-)\in\Ga^2\bigm|
\ga^+\cap\ga^-=\emptyset\Bigr\}\label{goodGa2}
\end{equation}
has full $\mu$-measure.
\end{proposition}
\begin{proof}
Take $\{\La_m\}_{m\in\N}\subset\B_c(\X)$ such that
$\bigcup_{m\in\N}\La_m=\X$. Then we can decompose the set
$\Ga^2\setminus\widetilde{\Ga}^2$ as
\[
\Ga^2\setminus\widetilde{\Ga}^2=\bigcup_{m\in\N} p^{-1}_{\La_m}
\bigl\{ (\ga^+,\ga^-)\in\Ga_{\La_m}^2\bigm|
\ga^+\cap\ga^-\neq\emptyset \bigr\},
\]
hence,
\[
\mu\bigl(\Ga^2\setminus\widetilde{\Ga}^2\bigr)\leq \sum_{m\in\N}
\mu^{\La_m} \Bigl(\bigl\{ (\ga^+,\ga^-)\in\Ga_{\La_m}^2\bigm|
\ga^+\cap\ga^-\neq\emptyset \bigr\}\Bigr).
\]
Since $\mu^{\La_m}$ is absolutely continuous w.r.t.
$\la_\sigma\times\la_\sigma$ it is enough to prove that
\[
(\la_\sigma\times\la_\sigma) \Bigl(\bigl\{
(\ga^+,\ga^-)\in\Ga_{\La_m}^2\bigm| \ga^+\cap\ga^-\neq\emptyset
\bigr\}\Bigr)=0.
\]
But if we denote for any fixed $\ga^+\in\Ga^+_{\La_m}$
\[
A_{\ga^+}:=\bigl\{ \ga^-\in\Ga^-_{\La_m}\bigm|
\ga^+\cap\ga^-\neq\emptyset \bigr\}
\]
then one has
\[
\la_\sigma (A_{\ga^+})\leq \sum_{x\in\ga^+}\la_\sigma\Bigl(\bigl\{
\ga^-\in\Ga^-_{\La_m}\bigm| x\in\ga^-\bigr\}\Bigr) =0.
\]
The remark that
\[
(\la_\sigma\times\la_\sigma) \Bigl(\bigl\{
(\ga^+,\ga^-)\in\Ga_{\La_m}^2\bigm| \ga^+\cap\ga^-\neq\emptyset
\bigr\}\Bigr)=\int_{\Ga^+_{\La_m}}A_{\ga^+}d\la_\sigma(\ga^+)
\]
is fulfilled the proof.
\end{proof}

\begin{proposition}\label{zeroset}
Let $\mu\in\M^1(\Ga^2)$ be a locally absolutely continuous measure
w.r.t. $\pi_\sigma\times\pi_\sigma$ and let $A$ be a
$\B(\X)$-measurable set such that $\sigma(A)=0$. Then the following
set
\[
B:=\bigl\{ (\ga^+,\ga^-)\in \Ga^2 \bigm| \ga^-\cap A\neq\emptyset
\bigr\}
\]
has zero $\mu$-measure.
\end{proposition}
\begin{proof}
Using the same trick as in the previous Proposition one can show
that it is enough to prove that for any $m\in\N$
\[
(\la_\sigma\times\la_\sigma) \Bigl(\bigl\{
(\ga^+,\ga^-)\in\Ga_{\La_m}^2\bigm| x\in A \text{ for some }
x\in\ga^- \bigr\}\Bigr)=0.
\]
But the left hand side is equal to
\begin{eqnarray*}
&&\la_\sigma(\Ga^+_{\La_m})\la_\sigma\Bigl(\bigl\{\ga^-\in\Ga^-_{\La_m}\bigm|
 x\in A \text{ for some }
x\in\ga^- \bigr\}\Bigr)\\& = & e^{\sigma(\La_m)}
\sum_{n=0}^\infty\frac{1}{n!}\sigma^\otimes\Bigl(\bigl\{
(x_1,\ldots,x_n)\in(\La_m)^n \bigm| x_i\in A \text{ for some }
i\bigr\}\Bigr)=0.
\end{eqnarray*}
The statement is proven.
\end{proof}

\begin{corollary}\label{forelementar}
Let $\mu\in\M^1(\Ga^2)$ be a locally absolutely continuous measure
w.r.t. $\pi_\sigma\times\pi_\sigma$. Then the set
  \[
    \bigl\{ (\ga^+,\ga^-,x)\in\Ga^2\times\X \bigm| x\in\ga^+ \bigr\}
  \]
  has $\mu\times\sigma$-measure $0$.
\end{corollary}

We define the {\em marginal distribution} of $\mu$ in a usual way,
namely,
\begin{equation}
d\mu^\pm(\ga^\pm):=\int_{\Ga^{\mp}}d\mu(\ga^+,\ga^-).\label{defmarg}
\end{equation}
Hence, for example, $\mu^+$ is a probability measure on
$\bigl(\Ga^+, \B(\Ga^+)\bigr)$. Then one can consider projection of
$\mu^+$ on $\Ga^+_{\La}$: $(\mu^+)^\La=\mu^+ \circ
\mathbf{p}_\La^{-1}$. On the other hand we may consider marginal
distribution of $\mu^\La$ whose we denote by $(\mu^\La)^+$.

It's easy to see that
\begin{equation}
(\mu^+)^\La=(\mu^\La)^+.\label{comuteprojmarg}
\end{equation}
Indeed, let $F:\Ga^2\rightarrow \R$ be a measurable function such
that there a exist measurable function $F^+:\Ga^+\rightarrow \R$
such that $F\left( \gamma ^{+},\gamma ^{-}\right) =F^{+}\left(
\gamma _{\Lambda }^{+}\right) $. Then
\begin{eqnarray*}
\int_{\Gamma ^2}F\left( \gamma ^{+},\gamma ^{-}\right) d\mu \left(
\gamma ^{+},\gamma ^{-}\right) &=&\int_{\Gamma _{\Lambda }^{+}\times
\Gamma _{\Lambda }^{-}}F\left( \gamma _{\Lambda }^{+},\gamma
_{\Lambda }^{-}\right) d\mu ^{\Lambda }\left( \gamma
_{\Lambda }^{+},\gamma _{\Lambda }^{-}\right)  \\
&=&\int_{\Gamma _{\Lambda }^{+}}F^{+}\left( \gamma _{\Lambda
}^{+}\right) \int_{\Gamma _{\Lambda }^{-}}d\mu ^{\Lambda }\left(
\gamma _{\Lambda }^{+},\gamma _{\Lambda }^{-}\right) =\int_{\Gamma
_{\Lambda }^{+}}F^{+}\left( \gamma _{\Lambda }^{+}\right) d\left(
\mu ^{\Lambda }\right) ^{+}\left( \gamma _{\Lambda }^{+}\right).
\end{eqnarray*}%
On the other hand
\begin{eqnarray*}
\int_{\Gamma ^2}F\left( \gamma ^{+},\gamma ^{-}\right) d\mu \left(
\gamma ^{+},\gamma ^{-}\right) &=&\int_{\Gamma ^{+}\times \Gamma
^{-}}F^{+}\left( \gamma ^{+}\right) d\mu
\left( \gamma ^{+},\gamma ^{-}\right)  \\
&=&\int_{\Gamma ^{+}}F^{+}\left( \gamma ^{+}\right) d\mu ^{+}\left(
\gamma ^{+}\right) =\int_{\Gamma _{\Lambda }^{+}}F^{+}\left( \gamma
_{\Lambda }^{+}\right) d\left( \mu ^{+}\right) ^{\Lambda }\left(
\gamma _{\Lambda }^{+}\right) .
\end{eqnarray*}%

\begin{remark}
Using \eqref{comuteprojmarg} it is clear that if $\mu$ is locally
absolutely continuous w.r.t. $\pi_\sigma\times\pi_\sigma$ then
$\mu^\pm$ are locally absolutely continuous w.r.t. $\pi_\sigma$.
\end{remark}

\begin{definition}\label{goodmeas}
We will say that locally absolutely continuous w.r.t.
$\pi_\sigma\times\pi_\sigma$ probability measure $\mu$ is satisfied
{\em local Ruelle bound} if for any $\La^\pm\in\B_c(\X)$ there exist
$C_{\La^\pm}>0$ such that for $\la_\sigma\times\la_\sigma$-a.a.
$(\eta^+,\eta^-)\in\Ga_{\La^+}^+\times\Ga_{\La^-}^-$
\begin{equation}\label{localRuellebound}
\frac{d\mu^{\La^+,\La^-}}{d(\la_\sigma\times\la_\sigma)}(\eta^+,\eta^-)\leq
(C_{\La^+})^{\vert\eta^+\vert}(C_{\La^-})^{\vert\eta^-\vert}.
\end{equation}
\end{definition}

For the measure $\mu$ from Definition~\ref{goodmeas} one can define
a~correlation function $k_\mu$, namely, for
$\la_\sigma\times\la_\sigma$-a.a.
$(\eta^+,\eta^-)\in\Ga^+_{\La^+}\times\Ga^-_{\La^-}$,
$\La^\pm\in\B_c(\X)$ we set
\begin{equation}\label{gencorfunc}
k_\mu(\eta^+,\eta^-)=\int_{\Ga^+_{\La^+}}\int_{\Ga^-_{\La^-}}
\frac{d\mu^{\La^+,\La^-}}{d(\la_\sigma\times\la_\sigma)}
(\eta^+\cup\xi^+,\eta^-\cup\xi^-)
d\la_\sigma(\xi^+)d\la_\sigma(\xi^-).
\end{equation}
Clearly,
\[
k_\mu(\emptyset,\emptyset)=1.
\]

It follows from infinitely-divisible property of $\la_\sigma$ that
r.h.s. of \eqref{gencorfunc} doesn't depend on $\La^\pm$. Also, from
definition of $\la_\sigma$ and \eqref{localRuellebound} one has that
\begin{equation}\label{localRuelleboundforcorfunct}
k_\mu(\eta^+,\eta^-)\leq
e^{C_{\La^+}\sigma(\La^+)}e^{C_{\La^-}\sigma(\La^-)}
(C_{\La^+})^{\vert\eta^+\vert}(C_{\La^-})^{\vert\eta^-\vert}.
\end{equation}

Correlation function of the marginal distribution $\mu^+$ we will
denote $k^+_\mu$ and define as
\begin{equation}\label{corfunc+}
k^+_\mu(\eta^+)=\int_{\Ga^+_{\La}}
\frac{d(\mu^+)^{\La}}{d\la_\sigma^\La} (\eta^+\cup\xi^+)
d\la_\sigma(\xi^+).
\end{equation}
for $\la_\sigma$-a.a. $\eta^+\in\Ga^+_{\La}$, $\La\in\B_c(\X)$.
Analogously, one can define $k^-_\mu$.

Putting in \eqref{gencorfunc} $\eta^-=\emptyset$, $\La^+=\La^-=\La$
we obtain using \eqref{comuteprojmarg}
\begin{eqnarray}
k_\mu(\eta^+,\emptyset)&=&\int_{\Ga^+_{\La}}\left(\int_{\Ga^-_{\La}}
\frac{d\mu^{\La}}{d(\la_\sigma\times\la_\sigma)}
(\eta^+\cup\xi^+,\xi^-)
d\la_\sigma(\xi^-)\right)d\la_\sigma(\xi^+)\notag\\
&=&\int_{\Ga^+_{\La}} \frac{d(\mu^{\La})^+}{d\la_\sigma}
(\eta^+\cup\xi^+) d\la_\sigma(\xi^+)=k^+_\mu(\eta^+).
\end{eqnarray}
Analogously,
\begin{equation}
k^-_\mu(\eta^-)=k_\mu(\emptyset,\eta^-).
\end{equation}

\section{Two-component Gibbs measures}

\begin{definition}\label{GibbsMeasure}
The measure $\mu \in \M^{1}\left( \Gamma^2 \right) $ is called a
{\em Gibbs measure} if there exist non-negative measurable functions
$r^{\pm }:\Gamma ^{2}\times X\rightarrow \left[ 0;+\infty \right) $
such that for all non-negative measurable functions $h_{1,2}:\Gamma
^{2}\times X\rightarrow \left[ 0;+\infty \right) $ the following
{\em partial Campbell---Mecke identities} hold
\begin{eqnarray}
&&\int_{\Gamma ^{2}}\sum_{x\in \gamma ^{+}}h_{1}\left( \gamma
^{+},\gamma ^{-},x\right) d\mu \left( \gamma ^{+},\gamma ^{-}\right)\label{CM+}\\
&=&\int_{\Gamma ^{2}}\int_{X}h_{1}\left( \gamma ^{+}\cup x,\gamma
^{-},x\right) r^{+}\left( \gamma ^{+},\gamma ^{-},x\right) d\sigma (
x ) d\mu \left( \gamma
^{+},\gamma ^{-}\right) , \notag \\
&&\int_{\Gamma ^{2}}\sum_{y\in \gamma ^{-}}h_{2}\left( \gamma
^{+},\gamma ^{-},y\right) d\mu \left( \gamma ^{+},\gamma ^{-}\right)\label{CM-}\\
&=&\int_{\Gamma ^{2}}\int_{X}h_{2}\left( \gamma ^{+},\gamma ^{-}\cup
y,y\right) r^{-}\left( \gamma ^{+},\gamma ^{-},y\right) d\sigma
\left( y\right) d\mu \left( \gamma ^{+},\gamma ^{-}\right) .\notag
\end{eqnarray}
\end{definition}

We denote class of such measures $\G(r^+,r^-,\sigma)$.

We will call the functions $r^\pm$ {\em partial
relative
energy densities} of the measure $\mu$.
With necessity these function have the following
properties.
\begin{lemma}
For $\mu $-a.a. $\left( \gamma ^{+},\gamma ^{-}\right) \in \Gamma
^{2}$ and for $\sigma $-a.a. $x,y\in X$ {\em the partial cocycle
identities}\ hold
\begin{eqnarray}
r^{+}\left( \gamma ^{+}\cup x^{\prime },\gamma ^{-},x\right)
r^{+}\left( \gamma ^{+},\gamma ^{-},x^{\prime }\right)
&=&r^{+}\left( \gamma ^{+}\cup x,\gamma ^{-},x^{\prime }\right)
r^{+}\left( \gamma ^{+},\gamma
^{-},x\right) , \label{CCI+} \\
r^{-}\left( \gamma ^{+},\gamma ^{-}\cup y^{\prime },y\right)
r^{-}\left( \gamma ^{+},\gamma ^{-},y^{\prime }\right)
&=&r^{-}\left( \gamma ^{+},\gamma ^{-}\cup y,y^{\prime }\right)
r^{-}\left( \gamma ^{+},\gamma ^{-},y\right) , \label{CCI-}
\end{eqnarray}
as well as {\em the balance identity} holds
\begin{equation}
r^{+}\left( \gamma ^{+},\gamma ^{-}\cup y,x\right) r^{-}\left(
\gamma ^{+},\gamma ^{-},y\right) =r^{-}\left( \gamma ^{+}\cup
x,\gamma ^{-},y\right) r^{+}\left( \gamma ^{+},\gamma ^{-},x\right)
.  \label{Bal}
\end{equation}
\end{lemma}

\begin{proof}
1. For any measurable $h_{1,2}:\Gamma \times X\rightarrow \left[ 0;+\infty
\right)$ we have using \eqref{CM+}
\begin{eqnarray*}
I &:=&\int_{\Gamma ^{2}}\sum_{x\in \gamma ^{+}}h_{1}\left( \gamma
^{+},\gamma ^{-},x\right) \sum_{x^{\prime }\in \gamma
^{+}}h_{2}\left( \gamma ^{+},\gamma ^{-},x^{\prime }\right) d\mu
\left( \gamma ^{+},\gamma
^{-}\right) \\
&=&\int_{\Gamma ^{2}}\int_{X}h_{1}\left( \gamma ^{+}\cup x,\gamma
^{-},x\right) \sum_{x^{\prime }\in \gamma ^{+}\cup x}h_{2}\left(
\gamma ^{+}\cup x,\gamma ^{-},x^{\prime }\right) r^{+}\left( \gamma
^{+},\gamma ^{-},x\right) d\sigma ( x ) d\mu \left( \gamma
^{+},\gamma
^{-}\right) \\
&=&\int_{\Gamma ^{2}}\int_{X}h_{1}\left( \gamma ^{+}\cup x,\gamma
^{-},x\right) \sum_{x^{\prime }\in \gamma ^{+}}h_{2}\left( \gamma
^{+}\cup x,\gamma ^{-},x^{\prime }\right) r^{+}\left( \gamma
^{+},\gamma ^{-},x\right) d\sigma ( x ) d\mu \left( \gamma
^{+},\gamma
^{-}\right) \\
&&+\int_{\Gamma ^{2}}\int_{X}h_{1}\left( \gamma ^{+}\cup x,\gamma
^{-},x\right) h_{2}\left( \gamma ^{+}\cup x,\gamma ^{-},x\right)
r^{+}\left( \gamma ^{+},\gamma ^{-},x\right) d\sigma ( x ) d\mu
\left( \gamma
^{+},\gamma ^{-}\right) \\
&=&\int_{\Gamma ^{2}}\int_{X}\int_{X}h_{1}\left( \gamma ^{+}\cup
x\cup x^{\prime },\gamma ^{-},x\right) h_{2}\left( \gamma ^{+}\cup
x\cup x^{\prime },\gamma ^{-},x^{\prime }\right) \\ &&\quad\times
r^{+}\left( \gamma ^{+}\cup x^{\prime },\gamma ^{-},x\right)
r^{+}\left( \gamma ^{+},\gamma ^{-},x^{\prime }\right) d\sigma
\left( x^{\prime }\right) d\sigma ( x ) d\mu \left( \gamma
^{+},\gamma ^{-}\right) \\
&&+\int_{\Gamma ^{2}}\int_{X}h_{1}\left( \gamma ^{+}\cup x,\gamma
^{-},x\right) h_{2}\left( \gamma ^{+}\cup x,\gamma ^{-},x\right)
r^{+}\left( \gamma ^{+},\gamma ^{-},x\right) d\sigma ( x ) d\mu
\left( \gamma ^{+},\gamma ^{-}\right),
\end{eqnarray*}%
and, analogously,
\begin{eqnarray*}
I&=&\int_{\Gamma ^{2}}\int_{X}\int_{X}h_{1}\left( \gamma ^{+}\cup
x\cup x^{\prime },\gamma ^{-},x\right) h_{2}\left( \gamma ^{+}\cup
x\cup x^{\prime },\gamma ^{-},x^{\prime }\right)\\ &&\qquad\times
r^{+}\left( \gamma ^{+}\cup x,\gamma ^{-},x^{\prime }\right)
r^{+}\left( \gamma ^{+},\gamma ^{-},x\right) d\sigma \left(
x^{\prime }\right) d\sigma ( x ) d\mu \left( \gamma ^{+},\gamma
^{-}\right)
\\
&+&\int_{\Gamma ^{2}}\int_{X}h_{1}\left( \gamma ^{+}\cup x,\gamma
^{-},x\right) h_{2}\left( \gamma ^{+}\cup x,\gamma ^{-},x\right)
 r^{+}\left( \gamma ^{+},\gamma ^{-},x\right) d\sigma ( x )
d\mu \left( \gamma ^{+},\gamma ^{-}\right) .
\end{eqnarray*}%
Comparing right hand sides of these equalities we obtain
(\ref{CCI+}). (\ref{CCI-}) is obtained in the same way.

2. For any measurable $h:\Gamma ^{2}\times X\times X\rightarrow \left[
0;+\infty \right) $ we have using \eqref{CM+}
and \eqref{CM-}
\begin{eqnarray*}
J &:=&\int_{\Gamma ^{2}}\sum_{x\in \gamma ^{+}}\sum_{y\in \gamma
^{-}}h\left( \gamma ^{+},\gamma ^{-},x,y\right) d\mu \left( \gamma
^{+},\gamma ^{-}\right) \\
&=&\int_{\Gamma ^{2}}\int_{X}\sum_{y\in \gamma ^{-}}h\left( \gamma
^{+}\cup x,\gamma ^{-},x,y\right) r^{+}\left( \gamma ^{+},\gamma
^{-},x\right)
d\sigma ( x ) d\mu \left( \gamma ^{+},\gamma ^{-}\right) \\
&=&\int_{\Gamma ^{2}}\int_{X}\int_{X}h\left( \gamma ^{+}\cup
x,\gamma ^{-}\cup y,x,y\right) r^{+}\left( \gamma ^{+},\gamma
^{-}\cup y,x\right) \\ &&\quad \times r^{-}\left( \gamma ^{+},\gamma
^{-},y\right) d\sigma \left( y\right) d\sigma ( x ) d\mu \left(
\gamma ^{+},\gamma ^{-}\right) ,
\end{eqnarray*}%
on the other hand,
\begin{equation*}
J=\int_{\Gamma ^{2}}\int_{X}\int_{X}h\left( \gamma ^{+}\cup x,\gamma
^{-}\cup y,x,y\right) r^{-}\left( \gamma ^{+}\cup x,\gamma
^{-},y\right) r^{+}\left( \gamma ^{+},\gamma ^{-},x\right) d\sigma
\left( y\right) d\sigma ( x ) d\mu \left( \gamma ^{+},\gamma
^{-}\right) .
\end{equation*}
Comparing right hand sides of these equalities we obtain
(\ref{Bal}).
\end{proof}

\begin{corollary}
As a result, we can define the {\em relative energy density} of the
measure $\mu$ as
\begin{equation}\label{ex:99}
r\left( \gamma ^{+},\gamma ^{-},x,y\right) :=r^{+}\left( \gamma
^{+},\gamma ^{-}\cup y,x\right) r^{-}\left( \gamma ^{+},\gamma
^{-},y\right)=r^{-}\left( \gamma ^{+}\cup x,\gamma ^{-},y\right)
r^{+}\left( \gamma ^{+},\gamma ^{-},x\right),
\end{equation}
and the following {\em Campbell---Mecke identity} holds%
\begin{multline}\label{CMfull}
\int_{\Gamma ^{2}}\sum_{x\in \gamma ^{+}}\sum_{y\in \gamma
^{-}}h\left( \gamma ^{+},\gamma ^{-},x,y\right) d\mu \left( \gamma
^{+},\gamma ^{-}\right) \\=\int_{\Gamma ^{2}}\int_{X}\int_{X}h\left(
\gamma ^{+}\cup x,\gamma ^{-}\cup y,x,y\right) r\left( \gamma
^{+},\gamma ^{-},x,y\right) d\sigma \left( y\right) d\sigma ( x )
d\mu \left( \gamma ^{+},\gamma ^{-}\right) .
\end{multline}
\end{corollary}

Next Lemma shows that the function $r$ also
satisfied cocycle identity.
\begin{lemma}
For $\mu $-a.a. $\left( \gamma ^{+},\gamma ^{-}\right) \in
\Gamma ^{2}$ and for $\sigma $-a.a. $x,x^{\prime },y,y^{\prime }\in X$%
\begin{equation}
r\left( \gamma ^{+}\cup x^{\prime },\gamma ^{-}\cup y^{\prime
},x,y\right) r\left( \gamma ^{+},\gamma ^{-},x^{\prime },y^{\prime
}\right) =r\left( \gamma ^{+}\cup x,\gamma ^{-}\cup y,x^{\prime
},y^{\prime }\right) r\left( \gamma ^{+},\gamma ^{-},x,y\right) .
\end{equation}
\end{lemma}

\begin{proof}
First of all let us prove that for $\mu $-a.a.
 $\left( \gamma ^{+},\gamma
^{-}\right) \in \Gamma ^{2}$ and for $\sigma $-a.a. $x,x^{\prime
},y,y^{\prime }\in X$%
\begin{eqnarray}
r^{+}\left( \gamma ^{+}\cup x,\gamma ^{-}\cup y,x^{\prime }\right)
r\left( \gamma ^{+},\gamma ^{-},x,y\right) &=&r\left( \gamma
^{+}\cup x,\gamma ^{-},x^{\prime },y\right) r^{+}\left(
\gamma ^{+},\gamma ^{-},x\right) \notag\\
&=&r\left( \gamma ^{+}\cup x^{\prime },\gamma ^{-},x,y\right)
r^{+}\left( \gamma ^{+},\gamma ^{-},x^{\prime }\right)\notag
\\ &=&r^{+}\left( \gamma ^{+}\cup x^{\prime },\gamma ^{-}\cup
y,x\right) r\left( \gamma ^{+},\gamma ^{-},x^{\prime },y\right) .
\label{ex:1}
\end{eqnarray}%
Really, using (\ref{Bal}), one has%
\begin{eqnarray*}
&& r^{+}\left( \gamma ^{+}\cup x,\gamma ^{-}\cup y,x^{\prime
}\right) r\left( \gamma ^{+},\gamma ^{-},x,y\right)
\\&=&r^{+}\left( \gamma ^{+}\cup x,\gamma ^{-}\cup y,x^{\prime
}\right) r^{+}\left( \gamma ^{+},\gamma ^{-}\cup y,x\right)
r^{-}\left( \gamma
^{+},\gamma ^{-},y\right)  \\
&=&r^{+}\left( \gamma ^{+}\cup x,\gamma ^{-}\cup y,x^{\prime
}\right) r^{-}\left( \gamma ^{+}\cup x,\gamma ^{-},y\right)
r^{+}\left( \gamma
^{+},\gamma ^{-},x\right) \\
&=&r\left( \gamma ^{+}\cup x,\gamma ^{-},x^{\prime },y\right)
r^{+}\left( \gamma ^{+},\gamma ^{-},x\right) ;
\end{eqnarray*}%
analogously,%
\[
r^{+}\left( \gamma ^{+}\cup x^{\prime },\gamma ^{-}\cup y,x\right) r\left(
\gamma ^{+},\gamma ^{-},x^{\prime },y\right) =r\left( \gamma ^{+}\cup
x^{\prime },\gamma ^{-},x,y\right) r^{+}\left( \gamma ^{+},\gamma
^{-},x^{\prime }\right) ;
\]
next, using (\ref{Bal}) and (\ref{CCI+}),%
we obtain
\begin{eqnarray*}
&&r\left( \gamma ^{+}\cup x^{\prime },\gamma ^{-},x,y\right)
r^{+}\left( \gamma ^{+},\gamma ^{-},x^{\prime }\right)\\
&=&r^{-}\left( \gamma ^{+}\cup x^{\prime }\cup x,\gamma
^{-},y\right) r^{+}\left( \gamma ^{+}\cup x^{\prime },\gamma
^{-},x\right)
r^{+}\left( \gamma ^{+},\gamma ^{-},x^{\prime }\right) \\
&=&r^{-}\left( \gamma ^{+}\cup x^{\prime }\cup x,\gamma
^{-},y\right) r^{+}\left( \gamma ^{+}\cup x,\gamma ^{-},x^{\prime
}\right) r^{+}\left(
\gamma ^{+},\gamma ^{-},x\right) \\
&=&r\left( \gamma ^{+}\cup x,\gamma ^{-},x^{\prime },y\right)
r^{+}\left( \gamma ^{+},\gamma ^{-},x\right) ,
\end{eqnarray*}%
that fulfilled \eqref{ex:1}.

In the same way we obtain that
\begin{eqnarray}
r^{-}\left( \gamma ^{+}\cup x,\gamma ^{-}\cup y,y^{\prime }\right)
r\left( \gamma ^{+},\gamma ^{-},x,y\right) &=&r\left( \gamma
^{+},\gamma ^{-}\cup y,x,y^{\prime }\right) r^{-}\left(
\gamma ^{+},\gamma ^{-},y\right) \notag\\
&=&r\left( \gamma ^{+},\gamma ^{-}\cup y^{\prime },x,y\right)
r^{-}\left(
\gamma ^{+},\gamma ^{-},y^{\prime }\right) \notag\\
&=&r^{-}\left( \gamma ^{+}\cup x,\gamma ^{-}\cup y^{\prime
},y\right) r\left( \gamma ^{+},\gamma ^{-},x,y^{\prime }\right) .
\label{ex:2}
\end{eqnarray}

As a result, using \eqref{Bal}, \eqref{ex:1}, \eqref{ex:2}, one has
\begin{eqnarray*}
&&r\left( \gamma ^{+}\cup x^{\prime },\gamma ^{-}\cup y^{\prime
},x,y\right)
r\left( \gamma ^{+},\gamma ^{-},x^{\prime },y^{\prime }\right) \\
&=&r^{-}\left( \gamma ^{+}\cup x^{\prime }\cup x,\gamma ^{-}\cup
y^{\prime },y\right) r^{+}\left( \gamma ^{+}\cup x^{\prime },\gamma
^{-}\cup y^{\prime
},x\right) r\left( \gamma ^{+},\gamma ^{-},x^{\prime },y^{\prime }\right) \\
&=&r^{-}\left( \gamma ^{+}\cup x^{\prime }\cup x,\gamma ^{-}\cup
y^{\prime },y\right) r\left( \gamma ^{+}\cup x,\gamma ^{-},x^{\prime
},y^{\prime
}\right) r^{+}\left( \gamma ^{+},\gamma ^{-},x\right) \\
&=&r\left( \gamma ^{+}\cup x,\gamma ^{-}\cup y,x^{\prime },y^{\prime
}\right) r^{-}\left( \gamma ^{+}\cup x,\gamma ^{-},y\right)
r^{+}\left(
\gamma ^{+},\gamma ^{-},x\right) \\
&=&r\left( \gamma ^{+}\cup x,\gamma ^{-}\cup y,x^{\prime },y^{\prime
}\right) r\left( \gamma ^{+},\gamma ^{-},x,y\right)
\end{eqnarray*}
that proves the statement.
\end{proof}

Cocycle and balance identities allow us to construct more complicate objects which characterized
energies between finite and infinite configurations.

\begin{definition}\label{def:2}
Let us fix some order of finite "+"-configuration $\eta ^{+}=\left\{
x_{1},x_{2},\ldots ,x_{n}\right\}$
and set%
\begin{eqnarray*}
R^{+}\left( \gamma ^{+},\gamma ^{-},\eta ^{+}\right) &=&R^{+}\left(
\gamma ^{+},\gamma ^{-},\left\{ x_{1},x_{2},\ldots ,x_{n}\right\}
\right) \\&:=& r^{+}\left( \gamma ^{+},\gamma ^{-},x_{1}\right)
r^{+}\left( \gamma ^{+}\cup x_{1},\gamma ^{-},x_{2}\right)
r^{+}\left( \gamma ^{+}\cup \left\{
x_{1},x_{2}\right\} ,\gamma ^{-},x_{3}\right)  \ldots \\
&&\times  r^{+}\left( \gamma ^{+}\cup \left\{ x_{1},x_{2},\ldots
,x_{n-1}\right\} ,\gamma ^{-},x_{n}\right) .
\end{eqnarray*}%
In \cite[Lemma 2.3]{FK05}, it was shown, in fact, that this
definition is correct (doesn't depend on the order of points in
$\eta ^{+}$) and moreover for any $\eta
_{1}^{+},\eta _{2}^{+}$ :%
\begin{equation}\label{ex:4}
R^{+}\left( \gamma ^{+},\gamma ^{-},\eta _{1}^{+}\cup \eta _{2}^{+}\right)
=R^{+}\left( \gamma ^{+},\gamma ^{-},\eta _{1}^{+}\right) R^{+}\left( \gamma
^{+}\cup \eta _{1}^{+},\gamma ^{-},\eta _{2}^{+}\right)
\end{equation}
(note that this fact doesn't depend on $\gamma ^{-}$). Let us set,
by definition,
\begin{equation}\label{empty+}
    R^+(\ga^+,\ga^-,\emptyset):=1.
\end{equation}
Note that \eqref{empty+} is consistent with \eqref{ex:4} if we put
there $\eta^+_1=\emptyset$.

In the same way we may define function $R^{-}\left( \gamma
^{+},\gamma ^{-},\eta ^{-}\right) $ fixing order $\eta ^{-}=\left\{
y_{1},y_{2},\ldots ,y_{m}\right\} $ and setting
\begin{eqnarray}
R^{-}\left( \gamma ^{+},\gamma ^{-},\eta ^{-}\right) &:=&r^{-}\left(
\gamma ^{+},\gamma ^{-},y_{1}\right) r^{-}\left( \gamma ^{+},\gamma
^{-}\cup y_{1},y_{2}\right) \ldots r^{-}\left( \gamma ^{+},\gamma
^{-}\cup \left\{ y_{1},\ldots ,y_{n-1}\right\} ,y_{n}\right)
,\notag\\
 R^-(\ga^+,\ga^-,\emptyset)&:=&1. \label{eq:yuuy}
\end{eqnarray}
And again%
\begin{equation}\label{ex:5}
R^{-}\left( \gamma ^{+},\gamma ^{-},\eta _{1}^{-}\cup \eta _{2}^{-}\right)
=R^{-}\left( \gamma ^{+},\gamma ^{-},\eta _{1}^{-}\right) R^{-}\left( \gamma
^{+}\cup \eta _{1}^{-},\gamma ^{-},\eta _{2}^{-}\right) .
\end{equation}
\end{definition}

Functions $R^\pm$ also satisfied balance
identities:

\begin{lemma}
For $\mu $-a.a. $\left( \gamma ^{+},\gamma ^{-}\right) \in \Gamma ^{2}$ and
for $\lambda _{\sigma }\times \lambda _{\sigma }$-a.a. $\left( \eta
^{+},\eta ^{-}\right) \in \Gamma _{0}^{2}$%
\begin{equation}\label{ex:d}
R^{+}\left( \gamma ^{+},\gamma ^{-}\cup \eta ^{-},\eta ^{+}\right)
R^{-}\left( \gamma ^{+},\gamma ^{-},\eta ^{-}\right) =R^{-}\left(
\gamma ^{+}\cup \eta ^{+},\gamma ^{-},\eta ^{-}\right) R^{+}\left(
\gamma ^{+},\gamma ^{-},\eta ^{+}\right) .
\end{equation}
\end{lemma}

\begin{proof}
Let $\left\vert \eta ^{-}\right\vert =1,
\eta^-=\{y\} $. Then we want to prove that%
\begin{equation}
R^{+}\left( \gamma ^{+},\gamma ^{-}\cup y,\eta ^{+}\right)
r^{-}\left( \gamma ^{+},\gamma ^{-},y\right) =r^{-}\left( \gamma
^{+}\cup \eta ^{+},\gamma ^{-},y\right) R^{+}\left( \gamma
^{+},\gamma ^{-},\eta ^{+}\right) . \label{ex:3}
\end{equation}
If $\left\vert \eta ^{+}\right\vert =1$ then \eqref{ex:3} holds due
to~\eqref{Bal}. Suppose that \eqref{ex:3} is true for any $\eta
^{+},$ such that $\left\vert \eta ^{+}\right\vert =n$. Then by
\eqref{ex:4}, \eqref{Bal}
\begin{eqnarray*}
&&R^{+}\left( \gamma ^{+},\gamma ^{-}\cup y,\eta ^{+}\cup x\right)
r^{-}\left( \gamma ^{+},\gamma ^{-},y\right) \\
&=&r^{+}\left( \gamma ^{+}\cup \eta ^{+},\gamma ^{-}\cup y,x\right)
R^{+}\left( \gamma ^{+},\gamma ^{-}\cup y,\eta ^{+}\right)
r^{-}\left(
\gamma ^{+},\gamma ^{-},y\right) \\
&=&r^{+}\left( \gamma ^{+}\cup \eta ^{+},\gamma ^{-}\cup y,x\right)
r^{-}\left( \gamma ^{+}\cup \eta ^{+},\gamma ^{-},y\right)
R^{+}\left(
\gamma ^{+},\gamma ^{-},\eta ^{+}\right) \\
&=&r^{-}\left( \gamma ^{+}\cup \eta ^{+}\cup x,\gamma ^{-},y\right)
r^{+}\left( \gamma ^{+}\cup \eta ^{+},\gamma ^{-},x\right)
R^{+}\left(
\gamma ^{+},\gamma ^{-},\eta ^{+}\right) \\
&=&r^{-}\left( \gamma ^{+}\cup \eta ^{+}\cup x,\gamma ^{-},y\right)
R^{+}\left( \gamma ^{+},\gamma ^{-},\eta ^{+}\cup x\right) ,
\end{eqnarray*}%
hence, \eqref{ex:3} holds.

Suppose now that we prove \eqref{ex:d} for any $\eta ^{-},$ s.t. $\left\vert \eta ^{-}\right\vert =n$ and consider
\begin{eqnarray*}
&&R^{+}\left( \gamma ^{+},\gamma ^{-}\cup \eta ^{-}\cup y,\eta
^{+}\right)
R^{-}\left( \gamma ^{+},\gamma ^{-},\eta ^{-}\cup y\right) \\
&=&R^{+}\left( \gamma ^{+},\gamma ^{-}\cup \eta ^{-}\cup y,\eta
^{+}\right) r^{-}\left( \gamma ^{+},\gamma ^{-}\cup \eta
^{-},y\right) R^{-}\left(
\gamma ^{+},\gamma ^{-},\eta ^{-}\right) \\
&=&r^{-}\left( \gamma ^{+}\cup \eta ^{+},\gamma ^{-}\cup \eta
^{-},y\right) R^{+}\left( \gamma ^{+},\gamma ^{-}\cup \eta ^{-},\eta
^{+}\right)
R^{-}\left( \gamma ^{+},\gamma ^{-},\eta ^{-}\right) \\
&=&r^{-}\left( \gamma ^{+}\cup \eta ^{+},\gamma ^{-}\cup \eta
^{-},y\right) R^{-}\left( \gamma ^{+}\cup \eta ^{+},\gamma ^{-},\eta
^{-}\right)
R^{+}\left( \gamma ^{+},\gamma ^{-},\eta ^{+}\right) \\
&=&R^{-}\left( \gamma ^{+}\cup \eta ^{+},\gamma ^{-},\eta ^{-}\cup
y\right) R^{+}\left( \gamma ^{+},\gamma ^{-},\eta ^{+}\right) .
\end{eqnarray*}
Hence, the statement of lemma is proved.
\end{proof}

\begin{corollary}
As a result, we can define
\begin{eqnarray}
R\left( \gamma ^{+},\gamma ^{-},\eta ^{+},\eta ^{-}\right)
&:=&R^{+}\left( \gamma ^{+},\gamma ^{-}\cup \eta ^{-},\eta
^{+}\right) R^{-}\left( \gamma ^{+},\gamma ^{-},\eta
^{-}\right)\notag
\\&=&R^{-}\left( \gamma ^{+}\cup \eta ^{+},\gamma ^{-},\eta
^{-}\right) R^{+}\left( \gamma ^{+},\gamma ^{-},\eta
^{+}\right).\label{ex:43}
\end{eqnarray}
\end{corollary}

Next statement is analog of properties \eqref{ex:4},
\eqref{ex:5} for the function $R$.

\begin{lemma}\label{le:spec}
For $\mu $-a.a. $\left( \gamma ^{+},\gamma ^{-}\right) \in \Gamma ^{2}$ and
for $\lambda _{\sigma }\times \lambda _{\sigma }$-a.a. $\left( \eta
_{1}^{+},\eta _{1}^{-}\right) ,\left( \eta _{2}^{+},\eta _{2}^{-}\right) \in
\Gamma _{0}^{2}$ the following equalities
hold
\begin{eqnarray*}
R\left( \gamma ^{+},\gamma ^{-},\eta _{1}^{+}\cup \eta _{2}^{+},\eta
^{-}\right) &=&R\left( \gamma ^{+}\cup \eta _{2}^{+},\gamma
^{-},\eta _{1}^{+},\eta ^{-}\right) R^{+}\left( \gamma ^{+},\gamma
^{-},\eta
_{2}^{+}\right) , \\
R\left( \gamma ^{+},\gamma ^{-},\eta ^{+},\eta _{1}^{-}\cup \eta
_{2}^{-}\right) &=&R\left( \gamma ^{+},\gamma ^{-}\cup \eta
_{2}^{-}{},\eta ^{+},\eta _{1}^{-}{}\right) R^{-}\left( \gamma
^{+},\gamma ^{-},\eta
_{2}^{-}{}\right) , \\
R\left( \gamma ^{+},\gamma ^{-},\eta _{1}^{+}\cup \eta _{2}^{+},\eta
_{1}^{-}\cup \eta _{2}^{-}\right) &=&R\left( \gamma ^{+}\cup \eta
_{2}^{+},\gamma ^{-}\cup \eta _{2}^{-}{},\eta _{1}^{+},\eta
_{1}^{-}{}\right) R\left( \gamma ^{+},\gamma ^{-},\eta _{2}^{+},\eta
_{2}^{-}\right).
\end{eqnarray*}
\end{lemma}

\begin{proof}
By \eqref{ex:43}, \eqref{ex:4} we obtain
\begin{eqnarray*}
R\left( \gamma ^{+},\gamma ^{-},\eta _{1}^{+}\cup \eta _{2}^{+},\eta
^{-}\right)  &=&R^{-}\left( \gamma ^{+}\cup \eta _{2}^{+}\cup \eta
_{1}^{+},\gamma ^{-},\eta ^{-}\right) R^{+}\left( \gamma ^{+},\gamma
^{-},\eta _{1}^{+}\cup
\eta _{2}^{+}\right) \\
&=&R^{-}\left( \gamma ^{+}\cup \eta _{2}^{+}\cup \eta
_{1}^{+},\gamma ^{-},\eta ^{-}\right) R^{+}\left( \gamma ^{+}\cup
\eta _{2}^{+},\gamma ^{-},\eta _{1}^{+}\right) R^{+}\left( \gamma
^{+},\gamma ^{-},\eta
_{2}^{+}\right) \\
&=&R\left( \gamma ^{+}\cup \eta _{2}^{+},\gamma ^{-},\eta
_{1}^{+},\eta ^{-}\right) R^{+}\left( \gamma ^{+},\gamma ^{-},\eta
_{2}^{+}\right) .
\end{eqnarray*}%
Second identity the may obtain in the same way.

Next, by first and second identities one has%
\begin{eqnarray*}
&&R\left( \gamma ^{+},\gamma ^{-},\eta _{1}^{+}\cup \eta
_{2}^{+},\eta _{1}^{-}\cup \eta _{2}^{-}\right)  \\ &=& R\left(
\gamma ^{+}\cup \eta _{2}^{+},\gamma ^{-},\eta _{1}^{+},\eta
_{1}^{-}\cup \eta _{2}^{-}\right) R^{+}\left( \gamma ^{+},\gamma
^{-},\eta
_{2}^{+}\right) \\
&=& R\left( \gamma ^{+}\cup \eta _{2}^{+},\gamma ^{-}\cup \eta
_{2}^{-}{},\eta _{1}^{+},\eta _{1}^{-}{}\right) R^{-}\left( \gamma
^{+}\cup \eta _{2}^{+},\gamma ^{-},\eta _{2}^{-}{}\right)
R^{+}\left( \gamma
^{+},\gamma ^{-},\eta _{2}^{+}\right) \\
&=& R\left( \gamma ^{+}\cup \eta _{2}^{+},\gamma ^{-}\cup \eta
_{2}^{-}{},\eta _{1}^{+},\eta _{1}^{-}{}\right) R\left( \gamma
^{+},\gamma ^{-},\eta _{2}^{+},\eta _{2}^{-}\right)
\end{eqnarray*}
that finished the proof.
\end{proof}

Next lemma shows that values of the function $R$ on some elements
may be defined directly via $r$.
\begin{lemma}\label{le:Rviar}
For $\la_\sigma\times\la_\sigma$-a.a. $(\eta ^{+},\eta
^{-})\in\Ga_0^2$ with $\left\vert \eta ^{+}\right\vert =\left\vert
\eta ^{-}\right\vert$ one has
\begin{eqnarray*}
R\left( \gamma ^{+},\gamma ^{-},\eta ^{+},\eta ^{-}\right)
&=&r\left( \gamma ^{+},\gamma ^{-},x_{1},y_{1}\right) r\left( \gamma
^{+}\cup x_{1},\gamma ^{-}\cup y_{1},x_{2},y_{2}\right)
\\&& \times  r\left( \gamma ^{+}\cup \left\{ x_{1},x_{2}\right\}
,\gamma
^{-}\cup \left\{ y_{1},y_{2}\right\} ,x_{3},y_{3}\right)  \ldots \\
&&\times   r\left( \gamma ^{+}\cup \left\{ x_{1},x_{2},\ldots
,x_{n-2}\right\} ,\gamma ^{-}\cup \left\{ y_{1},y_{2},\ldots
,y_{n-2}\right\}
,x_{n-1},y_{n-1}\right) \\
&& \times   r\left( \gamma ^{+}\cup \left\{ x_{1},x_{2},\ldots
,x_{n-1}\right\} ,\gamma ^{-}\cup \left\{ y_{1},y_{2},\ldots
,y_{n-1}\right\} ,x_{n},y_{n}\right)
\end{eqnarray*}%
for some fixed orders of points
\[
\eta ^{+}=\left\{ x_{1},x_{2},\ldots ,x_{n}\right\} ,~~~\eta ^{-}=\left\{
y_{1},y_{2},\ldots ,y_{n}\right\} .
\]
\end{lemma}

\begin{proof}
Let $\left\vert \eta ^{+}\right\vert =\left\vert \eta ^{-}\right\vert =1$, then the statement is followed from \eqref{ex:43}, Definition~\ref{def:2}
and \eqref{ex:99}.

Let us suppose that the statement is true for any $\eta ^{+},\eta ^{-},$ s.t. $%
\left\vert \eta ^{+}\right\vert =\left\vert \eta ^{-}\right\vert =n$. Then,
using \eqref{ex:43}, \eqref{ex:99}, \eqref{ex:3}
and Definition~\ref{def:2}, we obtain
\begin{eqnarray*}
&&r\left( \gamma ^{+}\cup \eta ^{+},\gamma ^{-}\cup \eta
^{-},x,y\right)
R\left( \gamma ^{+},\gamma ^{-},\eta ^{+},\eta ^{-}\right) \\
&=&r\left( \gamma ^{+}\cup \eta ^{+},\gamma ^{-}\cup \eta
^{-},x,y\right) R^{+}\left( \gamma ^{+},\gamma ^{-}\cup \eta
^{-},\eta ^{+}\right)
R^{-}\left( \gamma ^{+},\gamma ^{-},\eta ^{-}\right) \\
&=&r^{+}\left( \gamma ^{+}\cup \eta ^{+},\gamma ^{-}\cup \eta
^{-}\cup y,x\right) R^{-}\left( \gamma ^{+},\gamma ^{-},\eta
^{-}\right)  \\&&\quad\times r^{-}\left( \gamma ^{+}\cup \eta
^{+},\gamma ^{-}\cup \eta ^{-},y\right) R^{+}\left(
\gamma ^{+},\gamma ^{-}\cup \eta ^{-},\eta ^{+}\right) \\
&=&r^{+}\left( \gamma ^{+}\cup \eta ^{+},\gamma ^{-}\cup \eta
^{-}\cup y,x\right) R^{-}\left( \gamma ^{+},\gamma ^{-},\eta
^{-}\right)   \\&&\quad\times R^{+}\left( \gamma ^{+},\gamma
^{-}\cup \eta ^{-}\cup y,\eta ^{+}\right) r^{-}\left(
\gamma ^{+},\gamma ^{-}\cup \eta ^{-},y\right) \\
&=&R^{+}\left( \gamma ^{+},\gamma ^{-}\cup \eta ^{-}\cup y,\eta
^{+}\right) r^{+}\left( \gamma ^{+}\cup \eta ^{+},\gamma ^{-}\cup
\eta ^{-}\cup y,x\right)  \\&&\quad\times R^{-}\left( \gamma
^{+},\gamma ^{-},\eta ^{-}\right) r^{-}\left(
\gamma ^{+},\gamma ^{-}\cup \eta ^{-},y\right) \\
&=&R^{+}\left( \gamma ^{+},\gamma ^{-}\cup \eta ^{-}\cup y,\eta
^{+}\cup
x\right) R^{-}\left( \gamma ^{+},\gamma ^{-},\eta ^{-}\cup y\right) \\
&=&R\left( \gamma ^{+},\gamma ^{-},\eta ^{+}\cup x,\eta ^{-}\cup
y\right),
\end{eqnarray*}
that proves the assertion.
\end{proof}

Next theorem present Ruelle-type identity for Gibbs measure $\mu$
which also called ``infinitely divisible property''.
\begin{theorem} Let $\mu\in\G(r^+,r^-,\sigma)$. Then for any non-negative measurable
function $F:\Gamma ^{2}\rightarrow \left[
0;+\infty \right) $ and for any $\Lambda ^{\pm }\in\B_c(\X)$%
\begin{eqnarray}\notag
\int_{\Gamma ^{2}}F\left( \gamma \right) d\mu \left( \gamma
^{+},\gamma ^{-}\right) &=&\int_{\Gamma _{\Lambda
^{+}}^{+}}\int_{\Gamma _{\Lambda ^{-}}^{-}}\int_{\Gamma _{\Lambda
^{+c}}^{+}}\int_{\Gamma _{\Lambda ^{-c}}^{-}}F\left( \gamma ^{+}\cup
\eta ^{+},\gamma ^{-}\cup \eta ^{-}\right)\\[2mm] &&\times R\left( \gamma
^{+},\gamma ^{-},\eta ^{+},\eta ^{-}\right) d\mu \left( \gamma
^{+},\gamma ^{-}\right) d\lambda _{\sigma }\left( \eta ^{+}\right)
d\lambda _{\sigma }\left( \eta ^{-}\right).\label{Ruelle-full}
\end{eqnarray}
\end{theorem}
\begin{proof}
Set for $x\in\X$, $n\in\N$, $A^-\in\B(\Ga^-)$ and for measurable
non-negative measurable $F$
\[
h^{+}\left( \gamma ^{+},\gamma ^{-},x\right) =1_{A^{-}}\left( \gamma
^{-}\right) \1_{\left\{ \left\vert \gamma ^{+}\cap \Lambda
^{+}\right\vert =n\right\} }\1_{\Lambda ^{+}}( x ) F\left( \gamma
^{+},\gamma ^{-}\right).
\]%
Since
\[
\int_{\Gamma ^{2}}\sum_{x\in \gamma ^{+}}h^{+}\left( \gamma
^{+},\gamma
^{-},x\right) d\mu \left( \gamma ^{+},\gamma ^{-}\right) =n\int_{\Gamma ^{2}}%
\1_{\left\{ \left\vert \gamma ^{+}\cap \Lambda ^{+}\right\vert
=n\right\} }F\left( \gamma ^{+},\gamma ^{-}\right) 1_{A^{-}}\left(
\gamma ^{-}\right) d\mu \left( \gamma ^{+},\gamma ^{-}\right)
\]%
and
\begin{eqnarray*}
&&\int_{\Gamma ^{2}}\int_{X}h^{+}\left( \gamma ^{+}\cup x,x\right)
r^{+}\left( \gamma ^{+},\gamma ^{-},x\right) d\sigma ( x ) d\mu
\left( \gamma ^{+},\gamma ^{-}\right) \\
&=&\int_{\Gamma ^{2}}\int_{\Lambda ^{+}}\1_{\left\{ \left\vert
\left( \gamma ^{+}\cup x\right) \cap \Lambda ^{+}\right\vert
=n\right\} }1_{A^{-}}\left( \gamma ^{-}\right) F\left( \gamma
^{+}\cup x,\gamma ^{-}\right) r^{+}\left( \gamma ^{+},\gamma
^{-},x\right) d\sigma ( x ) d\mu \left( \gamma ^{+},\gamma ^{-}\right) \\
&=&\int_{\Gamma ^{2}}\int_{\Lambda ^{+}}\1_{\left\{ \left\vert
\gamma ^{+}\cap \Lambda ^{+}\right\vert =n-1\right\}
}1_{A^{-}}\left( \gamma ^{-}\right) F\left( \gamma ^{+}\cup x,\gamma
^{-}\right) r^{+}\left( \gamma ^{+},\gamma ^{-},x\right) d\sigma ( x
) d\mu \left( \gamma ^{+},\gamma ^{-}\right) ,
\end{eqnarray*}%
then using \eqref{CM+} we obtain
\begin{eqnarray*}
&&\int_{\Gamma ^{2}}\1_{\left\{ \left\vert \gamma ^{+}\cap \Lambda
^{+}\right\vert =n\right\} }1_{A^{-}}\left( \gamma ^{-}\right)
F\left(
\gamma ^{+},\gamma ^{-}\right) d\mu \left( \gamma ^{+},\gamma ^{-}\right) \\
&=&\frac{1}{n}\int_{\Lambda ^{+}}\int_{\Gamma ^{2}}\1_{\left\{
\left\vert \gamma ^{+}\cap \Lambda ^{+}\right\vert =n-1\right\}
}1_{A^{-}}\left( \gamma ^{-}\right) F\left( \gamma ^{+}\cup x,\gamma
^{-}\right) r^{+}\left( \gamma ^{+},\gamma ^{-},x\right) d\mu \left(
\gamma ^{+},\gamma ^{-}\right) d\sigma \left( y\right) d\sigma ( x )
\end{eqnarray*}%
for any non-negative measurable $F$. Apply this formula for function%
\[
\tilde{F}\left( \gamma ^{+},\gamma ^{-}\right) =F\left( \gamma
^{+}\cup x,\gamma ^{-}\right) r^{+}\left( \gamma ^{+},\gamma
^{-},x\right)
\]%
with fixed $x,y$. Then%
\begin{eqnarray*}
&&\int_{\Gamma ^{2}}\1_{\left\{ \left\vert \gamma ^{+}\cap \Lambda
^{+}\right\vert =n\right\} }1_{A^{-}}\left( \gamma ^{-}\right)
F\left(
\gamma ^{+},\gamma ^{-}\right) d\mu \left( \gamma ^{+},\gamma ^{-}\right) \\
&=&\frac{1}{n\left( n-1\right) }\int_{\Lambda
^{+}{}^{2}}\int_{\Gamma ^{2}}1_{A^{-}}\left( \gamma ^{-}\right)
\1_{\left\{ \left\vert \gamma ^{+}\cap \Lambda ^{+}\right\vert
=n-2\right\} }F\left( \gamma
^{+}\cup x_{1}\cup x_{2},\gamma ^{-}\right) \\
&&\quad\times r^{+}\left( \gamma ^{+}\cup x_{2},\gamma
^{-},x_{1}\right) r^{+}\left( \gamma ^{+},\gamma ^{-},x_{2}\right)
d\mu \left( \gamma ^{+},\gamma ^{-}\right) d\sigma \left(
x_{2}\right) d\sigma \left( x_{1}\right) .
\end{eqnarray*}%
Repeating this procedure we obtain, as a result,%
\begin{eqnarray*}
&&\int_{\Gamma ^{2}}\1_{\left\{ \left\vert \gamma ^{+}\cap \Lambda
^{+}\right\vert =n\right\} }1_{A^{-}}\left( \gamma ^{-}\right)
F\left(
\gamma ^{+},\gamma ^{-}\right) d\mu \left( \gamma ^{+},\gamma ^{-}\right) \\
&=&\frac{1}{n!}\int_{\Lambda ^{+}{}^{n}}\int_{\Gamma
^{2}}1_{A^{-}}\left( \gamma ^{-}\right) \1_{\left\{ \left\vert
\gamma ^{+}\cap \Lambda ^{+}\right\vert =0\right\} }F\left( \gamma
^{+}\cup \left\{ x_{1},\ldots
,x_{n}\right\} ,\gamma ^{-}\right) \\
&&\quad\times R^{+}\left( \gamma ^{+},\gamma ^{-},\left\{
x_{1},\ldots ,x_{n}\right\} \right) d\mu \left( \gamma ^{+},\gamma
^{-}\right) d\sigma \left( x_{1}\right) \ldots d\sigma \left(
x_{n}\right) .
\end{eqnarray*}%
Then%
\begin{eqnarray}
&&\int_{\Gamma ^{+}\times A^{-}}F\left( \gamma ^{+},\gamma
^{-}\right) d\mu
\left( \gamma ^{+},\gamma ^{-}\right) \notag\\
&=&\int_{\Gamma _{\Lambda ^{+}}^{+}}\int_{\Gamma _{\Lambda
^{+c}}^{+}}\int_{A^{-}}F\left( \gamma ^{+}\cup \eta ^{+},\gamma
^{-}\right) R^{+}\left( \gamma ^{+},\gamma ^{-},\eta ^{+}\right)
d\mu \left( \gamma ^{+},\gamma ^{-}\right) d\lambda _{\sigma }\left(
\eta ^{+}\right) \label{Ruelle+}.
\end{eqnarray}%
Analogously, for any $A^+\in\B(\Ga^+)$
\begin{eqnarray}
&&\int_{A^{+}\times \Ga^{-}}F\left( \gamma ^{+},\gamma ^{-}\right)
d\mu
\left( \gamma ^{+},\gamma ^{-}\right) \notag\\
&=&\int_{A^{+}}\int_{\Gamma _{\Lambda ^{-}}^{-}}\int_{\Gamma
_{\Lambda ^{-c}}^{-}}F\left( \gamma ^{+},\gamma ^{-}\cup \eta
^{-}\right) R^{-}\left( \gamma ^{+},\gamma ^{-},\eta ^{-}\right)
d\mu \left( \gamma ^{+},\gamma ^{-}\right) d\lambda _{\sigma }\left(
\eta ^{-}\right) \label{Ruelle-}.
\end{eqnarray}%

Putting $A^-=\Ga^-$ in \eqref{Ruelle+} and applying \eqref{Ruelle-}
to the r.h.s. of \eqref{Ruelle+} with $A^+=\Gamma _{\Lambda
^{+}}^{+}\times \Gamma _{\Lambda ^{+c}}^{+}$ we obtain
\begin{eqnarray*}
&&\int_{\Gamma ^{+}\times \Ga^{-}}F\left( \gamma ^{+},\gamma
^{-}\right) d\mu
\left( \gamma ^{+},\gamma ^{-}\right)\\
&=&\int_{\Gamma _{\Lambda ^{+}}^{+}}\int_{\Gamma _{\Lambda
^{+c}}^{+}}\int_{\Gamma _{\Lambda ^{-}}^{-}}\int_{\Gamma _{\Lambda
^{-c}}^{-}}F\left( \gamma ^{+}\cup \eta ^{+},\gamma ^{-}\cup \eta ^{-}\right)\\
&&\quad\times R^{+}\left( \gamma ^{+},\gamma ^{-}\cup \eta ^{-},\eta
^{+}\right) R^{-}\left( \gamma ^{+},\gamma ^{-},\eta ^{-}\right)d\mu
\left( \gamma ^{+},\gamma ^{-}\right) d\lambda _{\sigma }\left( \eta
^{+}\right)d\lambda _{\sigma }\left( \eta ^{-}\right).
\end{eqnarray*}
Hence, the statement is followed from \eqref{ex:43}.
\end{proof}

Next proposition shows that any Gibbs measure (in the sense of
Definition~\ref{GibbsMeasure}) is locally absolutely continuous
w.r.t. $\pi_\sigma\times\pi_\sigma$.
\begin{proposition}
Let $\mu\in\G(r^+,r^-,\sigma)$. Then for any $\La^\pm\in\B_c(\X)$
there exist
\begin{equation}\label{exppp}
\frac{d\mu^{\La^+,\La^-}}{d(\pi_\sigma^{\La^+}\times\pi_\sigma^{\La^-})}(\eta^+,\eta^-)
=e^{\sigma(\La^+)+\sigma(\La^-)}\int_{\Gamma _{\Lambda
^{+c}}^{+}}\int_{\Gamma _{\Lambda ^{-c}}^{-}} R\left( \gamma
^{+},\gamma ^{-},\eta ^{+},\eta ^{-}\right) d\mu \left( \gamma
^{+},\gamma ^{-}\right)
\end{equation}
for $\pi_\sigma^{\La^+}\times\pi_\sigma^{\La^-}$-a.a.
$(\eta^+,\eta^-)\in\Gamma _{\Lambda ^{+c}}^{+}\times\Gamma _{\Lambda
^{-c}}^{-}$.
\end{proposition}
\begin{proof}
For any measurable non-negative function $F$ such that
$F(\ga^+,\ga^-)=F(\ga^+_{\La^+},\ga^-_{\La^-})$,
by~\eqref{Ruelle-full}, we obtain
\begin{eqnarray*}
&&\int_{\Gamma _{\Lambda ^{+}}^{+}\times\Gamma _{\Lambda
^{-}}^{-}}F(\ga^+_{\La^+},\ga^-_{\La^-}) d\mu^{\La^+,\La^-} \left(
\ga^+_{\La^+},\ga^-_{\La^-}\right) = \int_{\Gamma ^{2}}F\left(
\gamma \right) d\mu \left( \gamma ^{+},\gamma ^{-}\right)\\
&=&\int_{\Gamma _{\Lambda ^{+}}^{+}}\int_{\Gamma _{\Lambda
^{-}}^{-}}F\left( \eta ^{+},\eta ^{-}\right)\int_{\Gamma _{\Lambda
^{+c}}^{+}}\int_{\Gamma _{\Lambda ^{-c}}^{-}}R\left( \gamma
^{+},\gamma ^{-},\eta ^{+},\eta ^{-}\right) d\mu \left( \gamma
^{+},\gamma ^{-}\right) d\lambda _{\sigma }\left( \eta ^{+}\right)
d\lambda _{\sigma }\left( \eta ^{-}\right),
\end{eqnarray*}
that fulfilled the statement.
\end{proof}

In particular for any $\mu\in\G(r^+,r^-,\sigma)$
Propositions~\ref{goodset} and \ref{zeroset} as well as
Corollary~\ref{forelementar} hold.

As we mentioned above, by \eqref{comuteprojmarg}, measure $\mu^+$ is
locally absolutely continuous w.r.t. $\pi_\sigma$ and for any
$A\in\B(\Ga^+_{\La})$, $\La\in\B_c(X)$
\[
(\mu^+)^{\La}(A)=(\mu^\La)^+(A)=\mu^\La(A\times{\Ga^-_\La}).
\]
Therefore, using \eqref{ex:43} and\eqref{Ruelle+}
\begin{eqnarray}
&&\frac{d\left( \mu ^{+}\right) ^{\Lambda }}{d\pi _{\sigma
}^{\Lambda }}\left( \eta
^{+}\right)\notag\\&=&e^{2\sigma(\La)}\int_{\Ga_\La^-}\int_{\Gamma
_{\Lambda ^{c}}^{+}}\int_{\Gamma _{\Lambda ^{c}}^{-}} R\left( \gamma
^{+},\gamma ^{-},\eta ^{+},\eta ^{-}\right) d\mu \left( \gamma
^{+},\gamma ^{-}\right)d\pi_\sigma^\La(\eta ^{-})\notag\\
&=&e^{\sigma(\La)}\int_{\Ga_\La^-}\int_{\Gamma _{\Lambda
^{c}}^{+}}\int_{\Gamma _{\Lambda ^{c}}^{-}} R^+\left( \gamma
^{+},\gamma ^{-}\cup\eta ^{-},\eta ^{+}\right) R^-\left( \gamma
^{+},\gamma ^{-},\eta ^{-}\right)d\mu \left( \gamma ^{+},\gamma
^{-}\right)d\la_\sigma^\La(\eta ^{-})\notag\\
&=&e^{\sigma(\La)}\int_{\Gamma _{\Lambda ^{c}}^{+}}\int_{\Gamma^{-}}
R^+\left( \gamma ^{+},\gamma ^{-},\eta ^{+}\right) d\mu \left(
\gamma ^{+},\gamma ^{-}\right)
\end{eqnarray}
for $\pi^{\La^+}_\sigma$-a.a. $\eta^+\in\Ga^+_{\La^+}$.

In the next proposition we find formulas for the correlation
functions of the Gibbs measures.
\begin{proposition}
Let $\mu\in\G(r^+,r^-,\sigma)$ and \eqref{localRuellebound} holds.
Then
\begin{eqnarray}
k_\mu(\eta^+,\eta^-)&=&\int_{\Gamma ^{2}}R( \gamma ^{+},\gamma
^{-},\eta ^{+},\eta^{-}) d\mu ( \gamma ^{+},\gamma
^{-}),\\
k^{+}_\mu( \eta ^{+})  &=&\int_{\Gamma ^{2}}R^{+}\left( \gamma
^{+},\gamma ^{-},\eta ^{+}\right) d\mu \left( \gamma ^{+},\gamma
^{-}\right).
\end{eqnarray}
\end{proposition}
\begin{proof}
Using \eqref{gencorfunc}, \eqref{exppp}, Lemma~\ref{le:spec} and
\eqref{Ruelle-full} we obtain
\begin{eqnarray*}
&&k_\mu(\eta^+,\eta^-)\\&=&\int_{\Ga^+_{\La^+}}\int_{\Ga^-_{\La^-}}
\int_{\Gamma _{\Lambda ^{+c}}^{+}}\int_{\Gamma _{\Lambda ^{-c}}^{-}}
R\left( \gamma ^{+},\gamma ^{-},
\eta^+\cup\xi^+,\eta^-\cup\xi^-\right) d\mu \left( \gamma
^{+},\gamma ^{-}\right)d\la_\sigma(\xi^+)d\la_\sigma(\xi^-)\\
&=&\int_{\Ga^+_{\La^+}}\int_{\Ga^-_{\La^-}} \int_{\Gamma _{\Lambda
^{+c}}^{+}}\int_{\Gamma _{\Lambda ^{-c}}^{-}} R\left( \gamma
^{+}\cup\xi^+,\gamma ^{-}\cup\xi^-, \eta^+,\eta^-\right) \\
&&\quad \times R\left( \gamma ^{+},\gamma ^{-}, \xi^+,\xi^-\right)
d\mu \left( \gamma
^{+},\gamma ^{-}\right)d\la_\sigma(\xi^+)d\la_\sigma(\xi^-)\\
&=&\int_{\Gamma ^{2}}R( \gamma ^{+},\gamma ^{-},\eta ^{+},\eta^{-})
d\mu ( \gamma ^{+},\gamma ^{-}).
\end{eqnarray*}
The second formula one can obtain in the same way or just putting
$\eta^-=\emptyset$ in the previous one and using \eqref{ex:43},
\eqref{eq:yuuy}.
\end{proof}

%

At the end of article we consider examples of partial relative
energies densities $r^\pm$ which satisfied
\eqref{CCI+}--\eqref{Bal}.

Let $\mu_{1,2}$ be Gibbs measures on $\bigl( \Ga, \B(\Ga) \bigl)$
with relative energies densities $r_{1,2}$ in the sense of
\cite{FK05}. Namely, let for any measurable
$h:\Ga\times\X\rightarrow [0; \infty)$
\[
\int_\Ga\sum_{x\in\ga}h(x,\ga)d\mu_{1,2}(\ga)=\int_\Ga\int_\X
h(x,\ga\cup x)r_{1,2}(\ga,x)d\sigma(x)d\mu_{1,2}(\ga).
\]
Let $\phi:\X^2\rightarrow\R\cup\{\infty\}$ be a symmetric function.
Then on can construct an example of $r^\pm$ which heuristically
corresponds to the following formal ``pair-potential perturbation''
$\mu\in\M^1(\Ga^2)$ of the product $\mu_1\times\mu_2$:
\[
d\mu(\ga^+,\ga^-)= " \frac{1}{Z}
\exp\Bigl\{-\sum\limits_{\{x,y\}\subset\ga}\phi(x,y)\Bigr\}d\mu_1(\ga^+)d\mu_2(\ga^-)".
\]
Namely, let
\[
r_0(\ga,x)=\exp\Bigl\{-\sum_{y\in\ga} \phi(x,y)\Bigr\},
\]
then one can set
\begin{eqnarray*}
r^+(\ga^+,\ga^-,x)&=&r_0(\ga^-,x)r_1(\ga^+,x),\\
r^-(\ga^+,\ga^-,y)&=&r_0(\ga^+,y)r_2(\ga^-,y).
\end{eqnarray*}
The partial cocycle identities \eqref{CCI+}, \eqref{CCI-} hold since
for $r_{1,2}$ the cocycle identities hold (see \cite{FK05}). One can
easily check the balance condition \eqref{Bal}:
\begin{eqnarray*}
r^+(\ga^+,\ga^-\cup y,x)r^-(\ga^+,\ga^-,y)&=& r_0(\ga^-\cup
y,x)r_1(\ga^+,x) r_0(\ga^+,y)r_2(\ga^-,y)\\
&=&e^{-\phi(x,y)} r_0(\ga^-,x)r_1(\ga^+,x) r_0(\ga^+,y)r_2(\ga^-,y)\\
&=&r_0(\ga^-,x)r_1(\ga^+,x) r_0(\ga^+\cup x,y)r_2(\ga^-,y) \\
&=&r^+(\ga^+,\ga^-,x) r^-(\ga^+\cup x,\ga^-,y).
\end{eqnarray*}

The simplest examples of $r_{1,2}$ are also pair potential
densities: let $\phi^\pm:\X^2\rightarrow\R\cup\{\infty\}$ be
symmetric functions and
\[
r_1(\ga^+,x)=\exp\Bigl\{-\sum_{x'\in\ga^+} \phi^+(x,x')\Bigr\},
\qquad r_2(\ga^-,y)=\exp\Bigl\{-\sum_{y'\in\ga^-}
\phi^-(y,y')\Bigr\}.
\]
Then $\mu_{1,2}$ are classical pair-potential Gibbs measures and
$\mu$ is a measure of type which is considered in \cite{GH}. As a
result, in this case
\begin{eqnarray*}
r^+(\ga^+,\ga^-,x)&=&\exp\Bigl\{-\sum_{y\in\ga^-}
\phi(x,y)-\sum_{x'\in\ga^+} \phi^+(x,x')\Bigr\},\\
r^-(\ga^+,\ga^-,y)&=&\exp\Bigl\{-\sum_{x\in\ga^+}
\phi(x,y)-\sum_{y'\in\ga^-} \phi^-(y,y')\Bigr\},
\end{eqnarray*}
and, therefore,
\begin{eqnarray*}
&&r(\ga^+,\ga^-,x,y)\\&=&\exp\Bigl\{-\phi(x,y) -\sum_{x'\in\ga^+}
\phi(y,x')-\sum_{y'\in\ga^-} \phi(x,y') - \sum_{x'\in\ga^+}
\phi^+(x,x') - \sum_{y'\in\ga^-} \phi^-(y,y') \Bigr\}.
\end{eqnarray*}

\end{document}